\documentclass[journal,twoside,web]{ieeecolor}
\usepackage{generic}
\usepackage{cite}
\usepackage{amsmath,amssymb,amsfonts}
\usepackage{algorithmic}
\usepackage{graphicx}
\usepackage{textcomp}

\usepackage{url}
\usepackage{bm}
\usepackage{amsmath}
\usepackage{amsfonts}
\usepackage{amssymb}
\usepackage{booktabs,multirow}
\usepackage{array}
\usepackage{subcaption}
\usepackage{lineno}
\usepackage[colorlinks=red, citecolor=blue]{hyperref}

\def\BibTeX{{\rm B\kern-.05em{\sc i\kern-.025em b}\kern-.08em
    T\kern-.1667em\lower.7ex\hbox{E}\kern-.125emX}}
\begin{document}
\title{Synthesizing MR Image Contrast Enhancement Using 3D High-resolution ConvNets}
\author{Chao Chen, Catalina Raymond, William Speier, Xinyu Jin, Timothy F. Cloughesy, Dieter Enzmann, Benjamin M. Ellingson, Corey W. Arnold
\thanks{This work was supported by the China Scholarship Council and NIH/NCI P50CA211015.}
\thanks{Chao Chen, and Xinyu Jin are with the Information Science and Electrical Engineering Department, Zhejiang Univeristy, 310027 Hangzhou, China. (e-mail: chench@zju.edu.cn)}
\thanks{Dieter Enzmann is in the Department of Radiological Sciences, David Geffen School of Medicine at the University of California Los Angeles.}
\thanks{Timothy F. Cloughesy is in the Department of Neurology, David Geffen School of Medicine at the University of California Los Angeles.}
\thanks{Catalina Raymond and B.M. Ellingson are in the UCLA Brain Tumor Imaging Laboratory, Center for Computer Vision and Imaging Biomarkers, Department of Radiological Sciences, David Geffen School of Medicine, University of California Los Angeles, Los Angeles.}
\thanks{W. Speier and C.W. Arnold are in the Computational Diagnostics Lab, the Department of Radiological Sciences, the Department of Pathology and Laboratory Medicine, and the Department of Bioengineering at the University of California Los Angeles, 924 Westwood Blvd, Suite 600, CA 90024 USA (Corresponding author: C.W. Arnold, e-mail: cwarnold@ucla.edu).}
}

\maketitle

\begin{abstract}
\textit{Objective:} Gadolinium-based contrast agents (GBCAs) have been widely used to better visualize disease in brain magnetic resonance imaging (MRI). However, gadolinium deposition within the brain and body has raised safety concerns about the use of GBCAs. Therefore, the development of novel approaches that can decrease or even eliminate GBCA exposure while providing similar contrast information would be of significant use clinically. \textit{Methods:} In this work, we present a deep learning based approach for contrast-enhanced T1 synthesis on brain tumor patients. A 3D high-resolution fully convolutional network (FCN), which maintains high resolution information through processing and aggregates multi-scale information in parallel, is designed to map pre-contrast MRI sequences to contrast-enhanced MRI sequences. Specifically, three pre-contrast MRI sequences, T1, T2 and apparent diffusion coefficient map (ADC), are utilized as inputs and the post-contrast T1 sequences are utilized as target output. To alleviate the data imbalance problem between normal tissues and the tumor regions, we introduce a local loss to improve the contribution of the tumor regions, which leads to better enhancement results on tumors. \textit{Results:} Extensive quantitative and visual assessments are performed, with our proposed model achieving a PSNR of 28.24dB in the brain and 21.2dB in tumor regions. \textit{Conclusion and Significance:} Our results suggest the potential of substituting GBCAs with synthetic contrast images generated via deep learning. Code is available at \url{https://github.com/chenchao666/Contrast-enhanced-MRI-Synthesis}
\end{abstract}


\begin{IEEEkeywords}
Medical Image Synthesis, GBCAs, Brain MRI, Contrast Enhancement, Fully Convolutional Networks.
\end{IEEEkeywords}

\begin{figure}[!t]
\begin{center}
\includegraphics[width=1.0\linewidth]{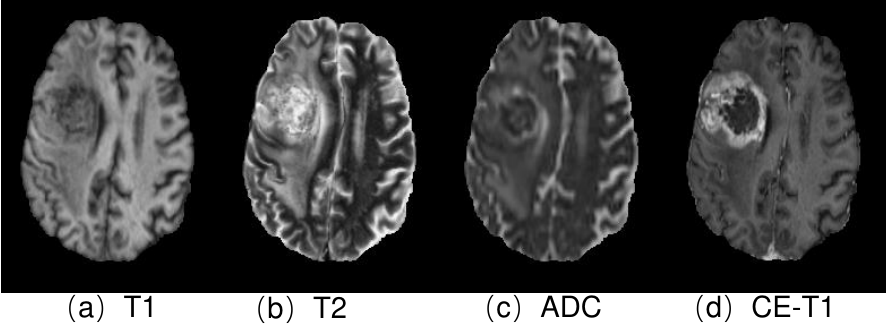}
\end{center}
\caption{Contrast and non-contrast MRI sequences used for brain tumor diagnosis and clinical monitoring. Three non-contrast scans, T1, T2, and the ADC were used to estimate the contrast-enhanced T1-weighted image (CE-T1) using a 3D FCN generator.}
\label{fig1}
\end{figure}

\section{Introduction}
Magnetic resonance imaging (MRI) is one of the most important techniques to distinguish different tissue properties and lesions in brain.  To better visualize different kinds of disease, gadolinium-based contrast agents (GBCAs) have been widely used for brain MRI image enhancement \cite{choi2019gadolinium}. Initially, the use of GBCAs was felt to carry minimal risk, with GBCAs administered in up to 35\% of all MRI examinations \cite{fraum2017gadolinium}. However, some recent studies have demonstrated the deposition of gadolinium contrast agents in body tissues, including the brain \cite{kanda2016gadolinium,gulani2017gadolinium}, which has raised broad safety concerns about the use of GBCAs in medical imaging. Previous studies have also suggested that GBCA dose should be as low as required, since advanced renal disease and the development of nephrogenic systemic fibrosis (NFS) are linked to high exposure to GBCAs \cite{khawaja2015revisiting,fraum2017gadolinium}. Even though deposition can be minimized by reducing the dose of gadolinium used, using low-dose contrast-enhanced MRI images may ignore some important information provided by contrast \cite{gong2018deep}. It is of importance to minimize or even eliminate the use of GBCAs, while preserving high-contrast information.

Recent development of deep learning methods have demonstrated success for medical image analysis \cite{litjens2017survey}, especially in the fields of segmentation\cite{ronneberger2015u,cciccek20163d,kamnitsas2017efficient,pereira2016brain}, detection \cite{shin2016deep,rajpurkar2017chexnet}, reconstruction, \cite{hammernik2018learning,mardani2018deep,he2022low} and synthesis \cite{bahrami2016convolutional,nie2018medical,wolterink2017deep,gong2018deep,dar2019image}. In this study, we focus on developing a deep learning based approach to synthesize contrast-enhanced brain MRI sequences from non-contrast brain MRI sequences. Specifically, synthetic contrast-enhanced MRI images would especially useful for certain patients, such as: (1) pediatric patients, (2) patients with benign or low grade (slow growing) brain tumors who are undergoing routine clinical exams over time to look at tumor growth, and (3) patients with impaired renal function or who can otherwise not get GBCAs.

Most recently, Gong et al. proposed to learn the reconstruction of full-dose T1 scans from pre-contrast T1 scans and 10\% low-dose T1 scans \cite{gong2018deep}. In order to completely eliminate the dependence on GBCAs, Kleesiek et al. proposed to predict contrast enhancement sequences directly from non-contrast brain MRI sequences \cite{kleesiek2019can}. To introduce additional information from other modalities, they utilized 10 multi-parametric scans as inputs. There are several limitations of existing studies. First, datasets used for training and evaluation are small, containing no more than 100 subjects. Second, off-the-shelf network architectures and loss functions are used, which likely limits performance. Finally, existing work has insufficient performance on tumors and small vessels. For these reasons, in this work, we introduce a larger scale dataset containing more than 400 MRI sequences, and design a 3D high resolution fully convolutional network (FCN) to synthesize contrast-enhanced T1 (CE-T1) images. Fig. \ref{fig1} illustrates the four MRI modalities used in this work. Specifically, T1, T2 and ADC, are used as inputs to synthesize the post-contrast T1 with the proposed 3D FCN model. The main contributions of this paper are:

\begin{itemize}
  \item A dataset of over 400 MRI sequences are analyzed, the largest explored thus far for the task of MRI virtual contrast enhancement.
  \item A 3D high-resolution FCN model is designed to generate the CE-T1 from the precontrast MRI scans. The presented model outperforms existing virtual contrast enhancement methods in two ways: (1) it maintains high-resolution information throughout processing, and (2) it repeats multi-scale fusion and aggregates multi-scale information in parallel.
  \item  Since the voxels that compose tumor regions are limited relative to the entire MRI volume, deep learning methods with global loss functions struggle to accurately synthesize contrast in these regions. Therefore, a local loss is introduced to re-balance the contribution of the tumor regions, which leads to improved performance on tumors.
  \item Extensive experiments, visual assessments, and ablation studies are conducted. As a result, we achieved a peak-signal-to-noise ratio (PSNR) of 28.24dB in brain regions and 21.2dB in tumor regions. Numerical and visual assessments demonstrate that the presented method significantly outperforms existing work.
\end{itemize}

\section{Related Work}
In this section, we review the deep network architectures and loss functions that are widely used for image-to-image translation. We then discuss recent applications in medical image synthesis that are related to our study.
\subsection{Image-to-Image Translation}
Image-to-image (I2I) translation has been explored in recent years with the aim of translating an input image in a source domain to an image in a target domain. The basic idea of I2I methods is to learn a non-linear feature mapping given the input and output image pairs as training data. A large number of network architectures has been proposed to act as the non-linear mapping. For example, Long et al. \cite{long2015fully} proposed to utilize a FCN model for image-to-segmentation. In Ronneberger et al. \cite{ronneberger2015u}, a U-Net architecture was proposed for biomedical image segmentation, which is currently widely used in medical image translation tasks. In Chen et al. \cite{chen2017deeplab}, the authors introduced the dilated convolution to enlarge the receptive field of neural networks. In Zhao et al. \cite{zhao2017pyramid}, a pyramid pooling module was proposed to fuse features under four different pyramid scales, which enable the model to utilize local and global context information for pixel-wise prediction. In Newell et al. \cite{newell2016stacked}, the stacked hourglass module was proposed to capture and consolidate information across all scales of the image for human pose estimation. Different from the traditional high-to-low and low-to-high FCN architectures, Sun et al. \cite{sun2019deep} proposed a high-resolution net, which maintains high-resolution feature maps throughout processing.

A large number of training objectives have been introduced to measure the difference between the generated image and the ground truth image in I2I translation tasks. The typical choices are $\ell_1$ and $\ell_2$ loss. In \cite{lai2017deep}, a differentiable variant of $\ell_1$ loss, named the Charbonnier penalty function, was proposed to handle outliers. Compared with $\ell_1$ or $\ell_2$ loss, which may lead to blurry images \cite{isola2017image}, adversarial loss \cite{isola2017image,han2018gan,dar2019image,nie2018medical} has become a popular choice for I2I tasks. This process trains a discriminator to distinguish generated images from ground truth images. Additionally, perceptual loss, which measures the difference in feature space, has also been widely used in I2I translation \cite{johnson2016perceptual,yang2018low}.

\begin{figure*}[!t]
\begin{center}
\includegraphics[width=1.0\linewidth]{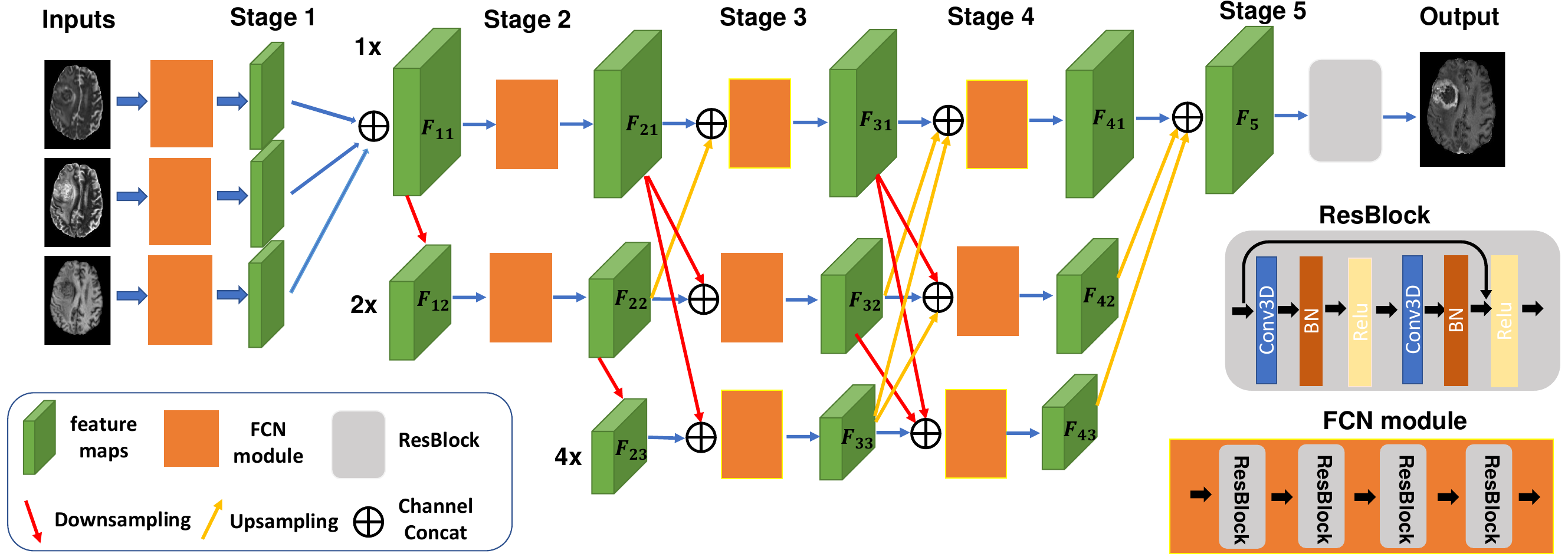}
\end{center}
\caption{Overview of our proposed model. A high-resolution FCN model was trained as a generator to synthesize contrast-enhanced T1 images. Three non-contrast scans, including T1, T2, and ADC, were utilized as input images.}
\label{fig2}
\end{figure*}

\subsection{Medical Image Synthesis}
Recently, an increasing number of machine learning and deep learning methods have shown potential in medical image synthesis, which estimates a desired imaging modality from other modalities or scans. For example, Li et al. applied the 3D-CNN model to predict missing PET patterns from MRI data \cite{li2014deep}. To improve the quality of 3T MR images, Bahrami et al. collected a dataset with paired 3T and 7T images scanned from the same subjects, and proposed to reconstruct 7T-like images from 3T images \cite{bahrami2016convolutional,bahrami2016reconstruction}. In Xiang et. al \cite{xiang2017deep}, the authors proposed to estimate standard-dose positron emission tomography (PET) images from low-dose PET and MRI images.  In Huang et al. \cite{huang2017simultaneous}, the authors proposed a weakly-supervised convolutional sparse coding method to simultaneously solve the problems of super-resolution and cross-modality image synthesis. In Dar et al. \cite{dar2019image}, the authors proposed a new approach for multi-contrast MRI synthesis based on conditional generative adversarial networks, employing adversarial loss to preserve intermediate-to-high frequency details. In Han et al. \cite{han2018gan}, a GAN model is employed to synthesize rich and diverse brain MR images from existing MR images. In Nie et al. \cite{nie2017medical,nie2018medical}, a 3D FCN model was trained to transform MRI to CT images using an adversarial strategy to train the FCN, which enforces the generated images to be more realistic. Finally, additional works have investigated methods for MRI to CT image synthesis \cite{xiang2018deep,emami2018generating,lei2019mri}, multimodal MRI synthesis \cite{chartsias2017multimodal,zhou2020hi}, and high-quality PET synthesis \cite{wang20183d}.

The studies that most relevant to ours can be seen in \cite{gong2018deep,kleesiek2019can,sun2020substituting}. Specifically, in \cite{gong2018deep}, the author utilized a U-Net-like model to synthesize full-dose CE-MRI from zero-dose pre-contrast MRI and the 10\% low-dose postcontrast MRI. In \cite{kleesiek2019can}, a 3D U-Net model was developed to generate CE-MRI, which utilizes 10 multiparametric MRI sequences acquired prior to GBCA application as inputs. As a result, their model demonstrates a peak signal to noise ratio (PSNR) of 22.967dB and a structural similarity index (SSIM) of 0.872 dB for the whole brain region. In \cite{sun2020substituting}, the author utilized the residual attention U-Net architecture to estimate CE-MRI from non-contrast T2 MRI for cerebral blood volume (CBV) mapping in mice brain.

\section{Materials and Methods}

\subsection{Data}
Our dataset was acquired at UCLA on Siemens 3 Tesla MRI systems as part of standard-of-care for brain tumor patients. The protocol used was consistent with the International Standard Brain Tumor Protocol \cite{ellingson2015consensus} and includes 3D MPRAGE T1-weighted pre- and post-contrast imaging, axial 2D T2-weighted imaging, and axial 2D diffusion-weighted imaging used in the calculation of the ADC map.  A total of 426 scans from 300 brain tumor patients were included in this study.  The data includes two parts, A and B. Set A consisted of 411 scans. It was used for training purposes and therefore further subdivided in 369 scans for training, and 42 scans randomly selected for validation.  Set B contained 15 test samples with precise tumor masks and were used to evaluate the quantitative performance on tumor regions. Note that scans in Set B are patients from the UCLA brain tumor trial (IRB\# 14-001261), and were selected randomly from the available data making sure both enhancing and non-enhancing tumor were part of the cohort. The experts with more than 10 years of experience created the tumor ROIs as part of the clinical trial reads.

Pre-contrast T1, T2 and ADC map, were utilized as input images and the contrast-enhanced T1 is utilized as the target image. Note that apparent diffusion coefficient (ADC) maps was chosen to augment T1 and T2 because it is independent of these image contrasts and may provide additional information for CE-T1 image synthesis. ADC maps were derived from standard, isotropic diffusion weighted images (DWIs) with and without diffusion weighting according to the standardized brain tumor imaging protocol (BTIP) \cite{ellingson2015jumpstarting}. Simply, ADC was calculated from b=0, 500, and 1000 s/mm2 by fitting the equation ADC=-1/b*ln(S(b)/S0), where b = 500 and 1,000 s/mm2, ln is the natural log, S(b) is the signal intensity for an MR image at the given b-value, and S0 is the signal intensity of the MR image without any diffusion weighting (b=0).

All the sequences of the MRI data were co-registered to match the targeted 3D contrast-enhanced T1. Bilinear interpolation was utilized to resize all the MRI data to the volume size of 192×256×192 voxels. To remove the skull,  brain masks were created for the 3D contrast-enhanced T1 sequences utilizing FSL’s  brain extraction tool \cite{jenkinson2005bet2}. Besides, in order to remove the side effects of the background slices, in the training and evaluation stage, we only selected the foreground slices and remove the top and bottom background slices which are less informative. Finally, all MRI scans were pre-processed by image equalization and the intensity values of the voxels within the brain region were normalized to [0,1].


\begin{figure*}[!t]
\begin{center}
\includegraphics[width=1.0\linewidth]{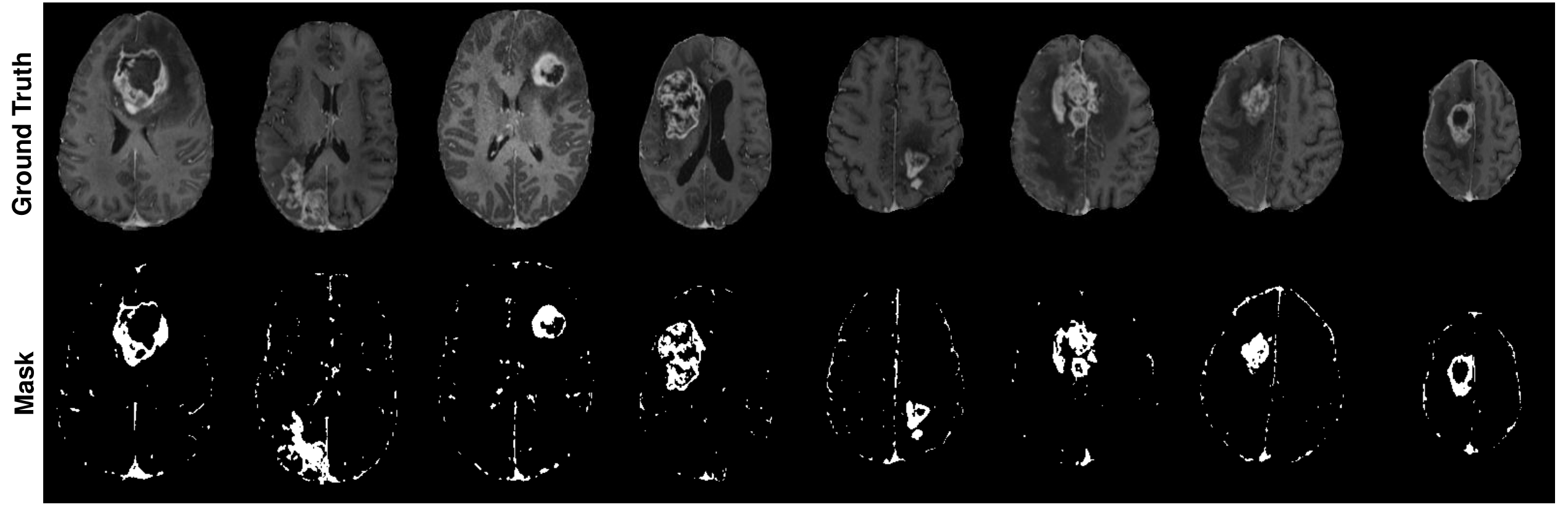}
\end{center}
\caption{Illustration of eight representative CE-T1 images and corresponding binary masks. The generated masks are able to identify the tumors, vessels and other high frequency details, which are enhanced by GBCAs.}
\label{fig3}
\end{figure*}

\subsection{Model Architecture}
Let $\mathbf{X}=[\mathbf{X}_{T1};\mathbf{X}_{T2};\mathbf{X}_{ADC}]\in\mathbb{R}^{h\times w\times d \times 3}$ denote the MRI sequences with three MRI scans, and $\mathbf{Y}\in\mathbb{R}^{h\times w\times d \times 1}$ denote the contrast-enhanced T1 sequence. To synthesize the CE-T1 sequences from the non-enhanced MRI scans, a 3D FCN model is designed to work as a non-linear mapping function $f_{\bm\theta}$, such that $f_{\bm\theta}: \mathbf{X}\rightarrow\mathbf{Y}$, where $\bm\theta$ is the model parameter to be learned.  As shown in Fig. \ref{fig2}, the introduced 3D FCN generator is composed of stacked 3D convolutional layers, batch normalization layers, and non-linear activation layers. The first stage is a stem network \cite{sun2019deep} which is composed of three individual FCN branches that used to handle different input modalities. The resulting feature maps corresponding to different modalities are then fused by  concatenation. Following the stem net are repeated multi-resolution subnetworks with generation multi-scale fusion stages. We start from a high-resolution subnetwork in the second stage and add high-to-low resolution subnetworks one-by-one gradually. Specifically, in the second stage, a $3\times 3\times 3$ convolution with stride 2 is used to obtain the $2\times$ downsampling feature maps. In the third stage, $4\times$ downsampling feature maps are obtained from the higher resolution feature maps. As a result, we have three different resolution feature maps for different subnetworks, which corresponds to multiple-scale information. To fuse the multi-resolution information comprehensively, multi-scale fusion stages are introduced, which ensure multi-resolution information exchange across different parallel subnetworks. Specifically, in the 3rd and 4th stages, different subnetworks aggregate the feature maps from the other parallel subnetworks. During the multi-scale fusion stages, upsample and downsample operators are utilized to match the size of the feature maps in different subnetworks. In the last stage, feature maps from different branches are fused by concatenation, and a ResBlock is utilized to obtain the final  CE-T1.

 As illustrated in Fig. \ref{fig2}, Let $F_{mn}$ denotes the feature maps generated by the $m$-th stage and the $n$-th subnetwork, then the feature maps can be calculated as: $F_{11} = \phi(T1)\oplus\phi(T2)\oplus\phi(ADC)$, $F_{12}=D_2(F_{11})$, $F_{21}=\phi(F_{11})$, $F_{22}=\phi(F_{12})$,  $F_{23}=D_2(F_{22})$, $F_{31}=\phi(F_{21}\oplus U_2(F_{22}))$, $F_{32}=\phi(F_{22}\oplus D_2(F_{21}))$, $F_{33}=\phi(F_{23}\oplus D_4(F_{21}))$, $F_{41}=\phi(F_{31}\oplus U_2(F_{32})\oplus U_4(F_{33}))$, $F_{42}=\phi(D_2(F_{31})\oplus F_{32}\oplus U_2(F_{33}))$, $F_{43}=\phi(D_4(F_{31})\oplus D_2(F_{32})\oplus F_{33})$ and $F_5=F_{41}\oplus U_2(F_{42})\oplus U_4(F_{43})$. Here, $\phi(x)$ denotes the feature mapping function determined by the FCN module parameters, $\oplus$ denotes feature map concatenation along channel dimension, $D_2$ denotes 2x downsampling, $D_4$ denotes 4x downsampling, $U_2$ denotes 2x upsampling and $U_4$ denotes 4x upsampling. The FCN module consists of four ResBlocks and each ResBlock is composed of two "Conv-BN-Relu" layers. The widths (number of feature maps) of the three parallel subnetworks are 64, 128, and 256. Detailed information regarding the model architecture can be seen in our source code, which is available at \url{https://github.com/chenchao666/Contrast-enhanced-MRI-Synthesis}.

The advantages of the presented model are four fold. First, we utilize three MRI scans as inputs and employ three individual stem nets for different modalities, which is able to preserve the modality-specific information. Second, the presented model maintains a high resolution representation throughout the processing pipeline, and contains three parallel subnetworks that can generate and process multi-scale information in parallel. Third, the repeated multi-scale fusion stages ensure better feature fusion across different scale. Fourth, 3D convolution was utilized to exploit additional information from neighboring slices.

\subsection{Loss Function}
Let $\mathbf{X}$ denote the input non-enhanced MRI sequences and $\mathbf{Y}$ denote the contrast-enhanced T1. Our goal is to learn a mapping function $f$ which can generate the CE-T1 sequences $\hat{\mathbf{Y}} = f_{\theta}(\mathbf{X})$, such that the synthetic CE-T1 is close to the ground truth $\mathbf{Y}$. The loss function utilized to train the model consists of three terms: pixel-wise MAE loss $\mathcal{L}_{MAE}$, Structural Similarity loss (SSIM) $\mathcal{L}_{SSIM}$ and a local loss $\mathcal{L}_{local}$ to focus performance on tumor regions.

$\bullet$ \textbf{Pixel-wise Loss:} The MAE loss and MSE loss are the most widely used pixel-wise losses for image synthesis. We found that using MSE loss resulted in blurrier images compared to MAE loss in this task. Therefore, the pixel-wise MAE loss was utilized in our model, which is given as
\begin{equation*}
\mathcal{L}_{MAE} = \Vert f_{\theta}(\mathbf{X})-\mathbf{Y}\Vert_1
\end{equation*}

$\bullet$ \textbf{SSIM Loss:} Using pixel-wise loss alone may ignore image structures. Therefore, we also utilize SSIM loss \cite{zhao2016loss}, which is perceptually motivated and leads to more realistic images. The SSIM loss is defined as
\begin{equation*}
\mathcal{L}_{SSIM} = \frac{1}{n}\sum_{i=1}^{n}\Vert1-SSIM(f_{\theta}(\mathbf{X})_i,\mathbf{Y}_i)\Vert_1
\end{equation*}
where $n$ denotes the number of slices of the output 3D MRI sequences, and $\mathbf{Y}_i$ denotes the $i$-th slice of the ground truth CE-T1. $SSIM(\mathbf{x},\mathbf{y})$ outputs a scalar between 0 and 1, which indicates the structural similarity between images $\mathbf{x}$ and $\mathbf{y}$. The definition of the SSIM metric can be seen in \ref{sec4}.

$\bullet$ \textbf{Local Loss:} Tumor regions are of particular interest, but account for a very small proportion of voxels in the entire MRI seqeuences. This data imbalance problem leads to under-fitting and poor performance for the tumor regions. Therefore, we introduce a local loss to increase the contribution of the tumor regions. The local loss is defined as
\begin{equation*}
\mathcal{L}_{local} = \Vert(f_{\bm\theta}(\mathbf{X})-\mathbf{Y})\odot\mathbf{M}\Vert_1
\end{equation*}
where $\mathbf{M}$ is a binary mask of the tumor regions, and $\odot$ denotes voxel-wise multiplication. Since it is very expensive to assign voxel-level labels for each slice, we do not have precise tumor masks for the training samples. Fortunately, compared to the non-enhanced T1 images, the tumors and vessels are significantly enhanced in the CE-T1 images due to the utilization of GBCAs. Therefore, we can calculate a rough tumor mask by thresholding the difference of the T1 and CE-T1 images as follows,
\begin{equation*}
\mathbf{M} =
\begin{cases}
1 & \mathbf{Y} - \mathbf{X}_{\text{T1}} > \delta \\
0 & \text{else}
\end{cases}
\end{equation*}
where $\delta$ is the threshold to control the size of the mask, $\mathbf{X}_{\text{T1}}$ is the non-enhanced T1w image, and $\mathbf{Y}$ is the contrast-enhanced T1 image. Fig. \ref{fig3} shows several examples of the generated binary masks. The generated masks are able to identify regions that are highly enhanced, such as tumors and vessels. These regions are used with our local loss to correct for data imbalance.

$\bullet$ \textbf{Overall Loss:} The overall loss function is defined as the weighted sum of all the three terms,
\begin{equation}
\mathcal{L}_{local} = \lambda_1\mathcal{L}_{MAE} + \lambda_2\mathcal{L}_{SSIM} + \lambda_3\mathcal{L}_{local}
\end{equation}
where $\lambda_1$, $\lambda_2$ and $\lambda_3$ are trade-off parameters to balance the contribution of each loss term.

\subsection{Implementation Details}
The proposed network was implemented in Python with the Keras library and trained on an NVIDIA DGX system with eight NVIDIA V100 CPUs and 512G memory. The Adam optimizer was utilized for training. Since feeding the whole 3D MRI sequences into the model leads to out-of-memory (OOM) problem, we follow \cite{nie2018medical,lei2019mri} to adopt a patch-based strategy for training, and set the batch size to three. In each iteration step, three-slices of MRI sequence with size $3\times256\times192$ are randomly sampled from each sequence volume. Therefore, we have three input channels with size $3\times3\times256\times192$ and one output with the size $3\times256\times192$. Model training is divided into two stages. In the first stage, we set $\lambda_1=1.0$, $\lambda_2=1.0$, and $\lambda_3=1.0$, and train the model for the first 40 epochs with a learning rate $lr = 0.0001$. In the second stage, we alter the trade-off parameter of the local loss by setting $\lambda_1=0.1$, $\lambda_2=0.1$, $\lambda_3=10$ and fine-tune the model for another 10 epochs with the learning rate $lr =0.00001$. We empirically the threshold utilized to obtain the brain mask to $\delta=0.1$ throughout the experiments.

\begin{figure*}[!t]
\begin{center}
\includegraphics[width=0.95\linewidth]{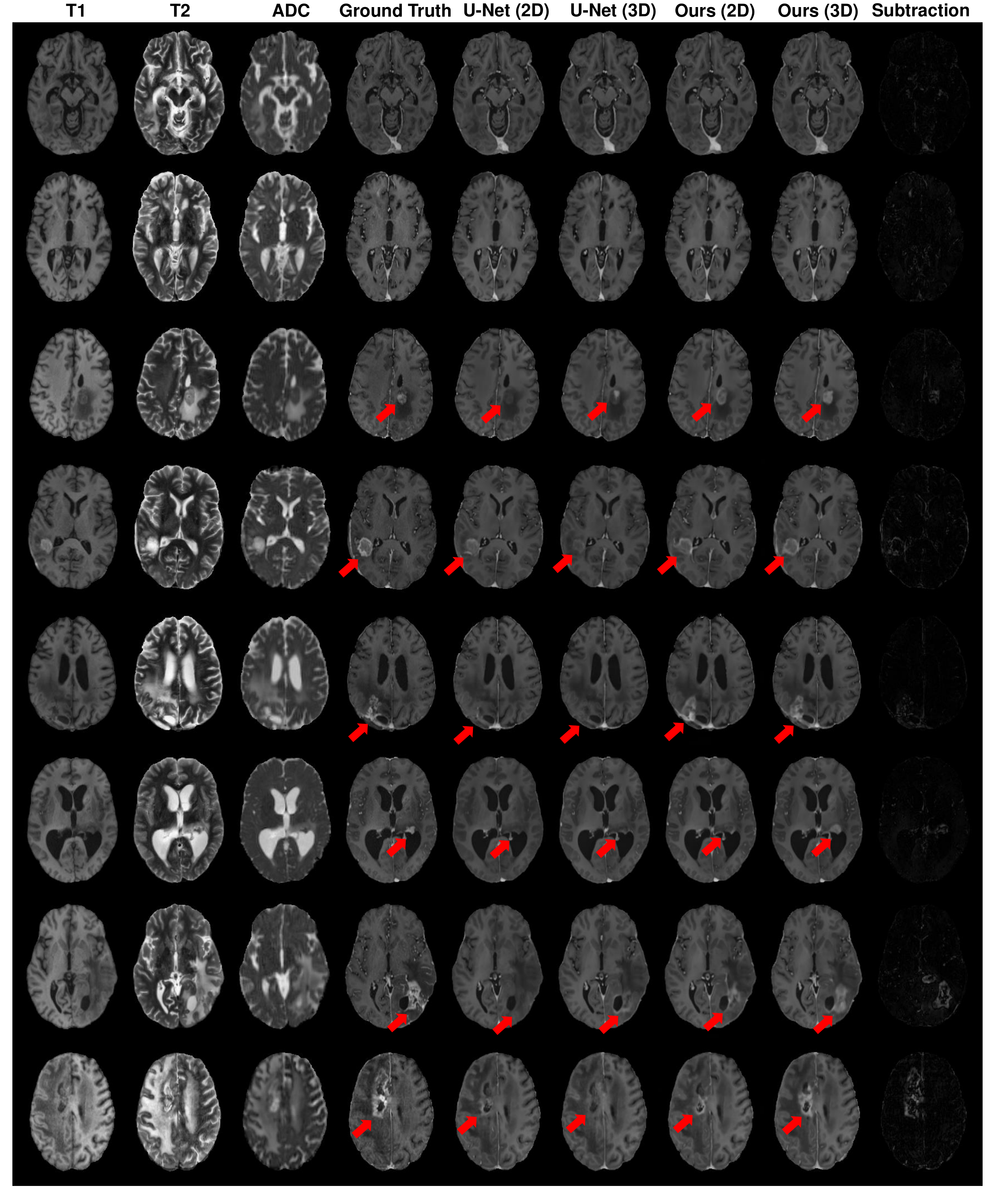}
\end{center}
\caption{Qualitative evaluation of our proposal and baseline models. From left to right, input T1, input T2, input ADC, ground truth image CE-T1, synthetic image of  a 2D U-Net,  synthetic image of a 3D U-Net, synthetic image of the proposed 2D FCN model, synthetic image of the proposed 3D FCN model, and the absolute difference between the results of our 3D FCN and the ground truth. The first two rows are from normal patients and the other rows are from patients with tumors. Different rows are from different subjects in the test set.}
\label{fig4}
\end{figure*}

\begin{figure*}[!t]
\begin{center}
\includegraphics[width=1.0\linewidth]{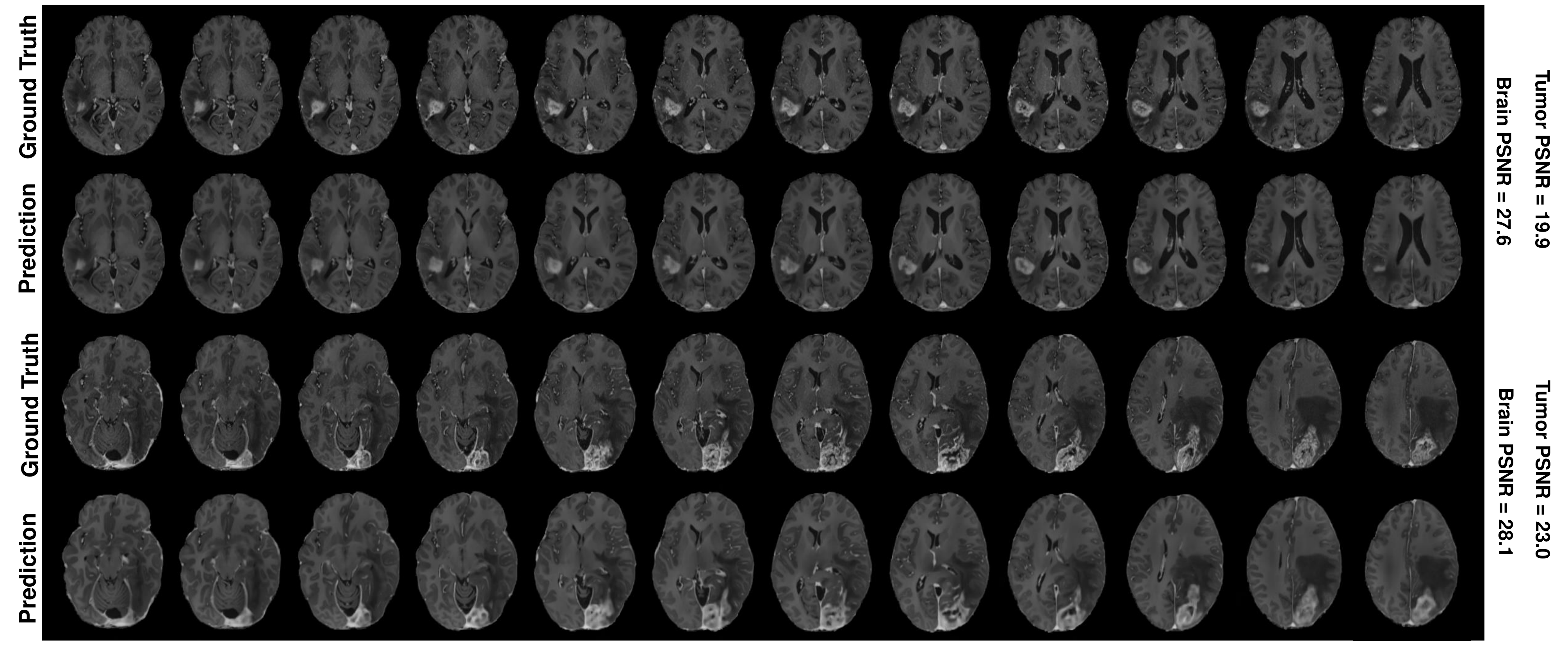}
\end{center}
\caption{Visual assessment of our proposed method in two representative test samples in set B. From top to bottom, the ground truth slices of test patient A, the synthetic CE-T1 for test patient A, the ground truth slices of test patient B, the synthetic CE-T1 for test patient B. The synthetic CE-T1 images are generated by the proposed 3D FCN model. The images in the same row represent different slices of the same subject.}
\label{fig5}
\end{figure*}

\section{Experiments}
In this section, we first introduce the baseline models and evaluation metrics that we utilized in the experiments. Then, the qualitative and quantitative performance of different models are presented. Finally, we provide ablation experiments that show the impact of different input MRI scans and the impact of the introduced local loss.
\subsection{Baseline Model and Evaluation Metric}
\label{sec4}
$\bullet$ \textbf{Baseline Model:} To evaluate the effectiveness of our proposed model, we implemented a 2D U-Net, a 3D U-Net, as well as the 2D version of our proposed network. All the baseline models are evaluated on the same training and test sets. For the 2D U-Net, we expanded the model depth to 2x of the original U-Net network. For the 3D U-Net, we expanded the model depth to 1.5x of the original model depth.

$\bullet$ \textbf{Evaluation Metric:} The quantitative performance of our model and the baseline models was measured using mean absolute error (MAE), peak signal-to-noise ratio (PSNR) and structural similarity index (SSIM). Given two images $\mathbf{y}\in\mathbf{Y}$ and $\hat{\mathbf{y}}\in\hat{\mathbf{Y}}$, where $\mathbf{y}$ is the ground truth image and $\hat{\mathbf{y}}$ is the predicted image. Then,
\begin{itemize}
  \item $ MAE(\mathbf{Y},\hat{\mathbf{Y}}) = \frac{1}{\Omega_{\mathbf{Y}}}\Vert\mathbf{Y}-\hat{\mathbf{Y}}\Vert_1 $
  \item $ MSE(\mathbf{Y},\hat{\mathbf{Y}}) = \frac{1}{\Omega_{\mathbf{Y}}}\Vert\mathbf{Y}-\hat{\mathbf{Y}}\Vert_2^2 $
  \item $ PSNR(\mathbf{Y},\hat{\mathbf{Y}}) = 10\log_{10}(\tfrac{MAX_I^2}{MSE})$
  \item $SSIM(\mathbf{Y},\hat{\mathbf{Y}}) = \frac{1}{\Phi_{\mathbf{Y}}}\sum\limits_{\mathbf{y},\hat{\mathbf{y}}} \frac{(2\mu_{\mathbf{y}}\mu_{\hat{\mathbf{y}}}+C1)(2\sigma_{\mathbf{y}\hat{\mathbf{y}}}+C2)}{(\mu_{\mathbf{y}}^2+\mu_{\hat{\mathbf{y}}}^2+C1)(\sigma_{\mathbf{y}}^2+\sigma_{\hat{\mathbf{y}}}^2+C2)}$
\end{itemize}
where $\Omega_{\mathbf{Y}}$ is the number of voxels in $\mathbf{Y}$, $\Phi_{\mathbf{Y}}$ is the number of slices in $\mathbf{Y}$, $\mu_{\mathbf{y}}$ and $\sigma_{\mathbf{y}}$ are the mean and variance value of image $\mathbf{y}$, and $\sigma_{\mathbf{y}\hat{\mathbf{y}}}$ is the covariance between the ground truth image $\mathbf{y}$ and the predicted image $\hat{\mathbf{y}}$. $MAX_I$ is the maximum value of the image $\mathbf{y}$.  Theoretically, lower MAE values and higher PSNR and SSIM values indicate better image generation quality. The statistical significance of experimental results was evaluated using paired t-tests.

\subsection{Experimental Results}
\textbf{Qualitative and Visual Assessment} The results of representative test samples are shown in Fig. \ref{fig4}. The first two rows demonstrate a normal subject without any tumor and the remaining six rows demonstrate subjects with tumors. We compared the visual performance between our model (both 2D and 3D model) and the 2D/3D U-Net, which are widely used techniques for medical image synthesis tasks \cite{gong2018deep,kleesiek2019can}.  As can be seen, the results reveal several interesting observations. First, both U-Net models and our proposed model generate promising visual performance for the normal patients. The vessels and high-frequency details in the generated images are very close to the ground truth images. Second, since the tumor voxels are limited relative to the entire MRI volume, models are more likely to over-fit to normal regions. As a result, the performance of the tumor regions is much worse than the performance on the normal regions. Third, compared with the U-Net model, which misses or under-estimates most of the tumors for the abnormal samples, our model often yields better visual performance for the tumor regions. Fourth, compared with the 2D model, the 3D model achieves better performance, especially for tumor regions. We believe this is because the 3D model takes advantage of the information from the neighboring slices. Fifth, while achieving promising visual performance for the whole brain MRI, our model sometimes misses or under-estimates some tumors, especially for those that are not distinct enough in the non-contrast images. Finally, Fig. \ref{fig5} shows two representative MRI sequences in test set B. As observed, our model synthesized satisfactory contrast enhanced T1 images for different slices of the given 3D volume.

\begin{table}[]
\centering
\caption{Quantitative comparison between our model and baseline models. The models are evaluated on test set A, and the performance on the whole brain region is presented.}
\label{tab1}
\begin{tabular}{|c|c|c|c|}
\hline
\textbf{Model}     & \textbf{MAE} & \textbf{PSNR} & \textbf{SSIM} \\ \hline
\textbf{Ref} \cite{kleesiek2019can}      & N/A   & 22.97$\pm$1.16      &  0.872$\pm$0.031    \\ \hline
\textbf{U-Net} (2D) &  0.033$\pm$0.005   &   26.86$\pm$1.05   &  0.905$\pm$0.038     \\ \hline
\textbf{U-Net} (3D) &  0.032$\pm$0.004   &   27.21$\pm$1.18   &  0.908$\pm$0.038    \\ \hline
\textbf{Ours} (2D)  &  0.030$\pm$0.005   &   27.87$\pm$1.30   &  0.915$\pm$0.039    \\ \hline
\textbf{Ours} (3D)  &  \textbf{0.029$\pm$0.005}   &   \textbf{28.24$\pm$1.26}   &  \textbf{0.923$\pm$0.041}    \\ \hline
\end{tabular}
\end{table}

\textbf{Quantitative Evaluation}
As shown in Table \ref{tab1}, we show the quantitative performance of our proposal and comparison methods in test set A. All metrics are computed based on the brain region. Our model significantly outperforms the U-Net model in PSNR and SSIM metrics and the 3D FCN model outperforms its 2D counterpart, which is consistent with our visual assessment. Specifically, U-Net(3D) outperforms U-Net(2D) by 0.35 points in PSRN (p-value=0.021), and Ours(3D) outperforms Our(2D) by 0.37 points (p-value=0.018), which demonstrates the effectiveness of utilizing 3D spatial information. Besides, the proposed Ours(3D) significantly outperforms the U-Net(3D) by more than one point (P-value=0.00037), we believe this is because our proposed model maintains a high resolution representation throughout the processing pipeline and contains three parallel subnetworks with multi-scale fusion stages. Compared with an existing method \cite{kleesiek2019can} that performs virtual contrast enhancement with deep learning, our proposal has more than five points improvement in PSNR, and has five points improvement in SSIM. Note that in \cite{kleesiek2019can}, the author utilized 10-channel multiparametric MRI data as input while we only utilize three, which is a subset of their data. Our model outperforms \cite{kleesiek2019can} by a large margin even with less input data, which demonstrates the superiority of our proposed framework.

In order to evaluate the quantitative performance on the tumor region, we also collected 15 test patients with precise tumor masks in set B. The test performance in set B is shown in Table \ref{tab2}, with performance on both brain region and tumor region presented. The overall performance on the brain region is similar to the results on set A. Our proposed method significantly outperforms the U-Net and existing work \cite{kleesiek2019can}. For results on tumors, the proposed model outperforms the U-Net(3D) model by a large margin (p=0.0036), and the best performance on tumors is 21.2 in PSNR. Note that the quantitative performance on tumors is far from perfect and much worse than the performance on the whole brain region, this is because the tumor pixels are out-of-distribution and the model therefore tends to underestimate on tumor regions. It is worth noting that \cite{kleesiek2019can} utilizes a U-Net shape model to segment the tumor masks, which they utilize to evaluate the performance on tumors.
These masks were not reviewed by a radiologist and therefore their quantitative performance on tumors may be inaccurate due to segmentation errors.

\begin{table}[]
\centering
\caption{Quantitative performance evaluated on test set B. The performance on the brain region and tumor region are presented.}
\label{tab2}
\begin{tabular}{|c|c|c|c|}
\hline
\multirow{2}{*}{\textbf{Model}} & \multicolumn{2}{c|}{\textbf{PSNR}} & \textbf{SSIM}  \\ \cline{2-4}
                       & \textbf{Brain}       & \textbf{Tumor}       & \textbf{Brain} \\ \hline
\textbf{Ref \cite{kleesiek2019can}}                    & 22.97$\pm$1.16             & 20.15$\pm$4.70            & 0.872$\pm$0.031        \\ \hline
\textbf{UNet (2D)}              & 26.44$\pm$1.40             &18.45$\pm$2.22             & 0.896$\pm$0.022      \\ \hline
\textbf{UNet (3D)}              & 26.79$\pm$1.26             &18.89$\pm$2.38             & 0.899$\pm$0.022      \\ \hline
\textbf{Ours (2D)}              & 27.22$\pm$1.21             &19.64$\pm$2.59             & 0.903$\pm$0.023       \\ \hline
\textbf{Ours (3D)}              & \textbf{27.62$\pm$1.34}             &\textbf{21.2$\pm$2.36}              &\textbf{0.909$\pm$0.023}       \\ \hline
\end{tabular}
\end{table}

\begin{table}[]
\centering
\caption{Comparison experiments by using different modalities as inputs.}
\label{tab3}
\begin{tabular}{|c|c|c|c|c|}
\hline
\multicolumn{2}{|c|}{\textbf{Input}}     & \textbf{T1}          & \textbf{T1+T2}       & \textbf{T1+T2+ADC}   \\ \hline
\multirow{2}{*}{\textbf{A-Brain}} & \textbf{PSNR} & 27.4$\pm$1.28   & 27.9$\pm$1.24   & \textbf{28.24$\pm$1.26}  \\ \cline{2-5}
                         & \textbf{SSIM} & 0.914$\pm$0.040 & 0.920$\pm$0.040 & \textbf{0.923$\pm$0.041} \\ \hline
\multirow{2}{*}{\textbf{B-Brain}} & \textbf{PSNR} & 26.7$\pm$1.23   & 27.3$\pm$1.27   & \textbf{27.62$\pm$1.34}  \\ \cline{2-5}
                         & \textbf{SSIM} & 0.898$\pm$0.022 & 0.905$\pm$0.023 & \textbf{0.909$\pm$0.023} \\ \hline
\textbf{B-Tumor}                  & \textbf{PSNR} & 19.8$\pm$2.32   & 20.8$\pm$2.30   & \textbf{21.2$\pm$2.36}   \\ \hline
\end{tabular}
\end{table}

\begin{figure}[!t]
\begin{center}
\includegraphics[width=1.0\linewidth]{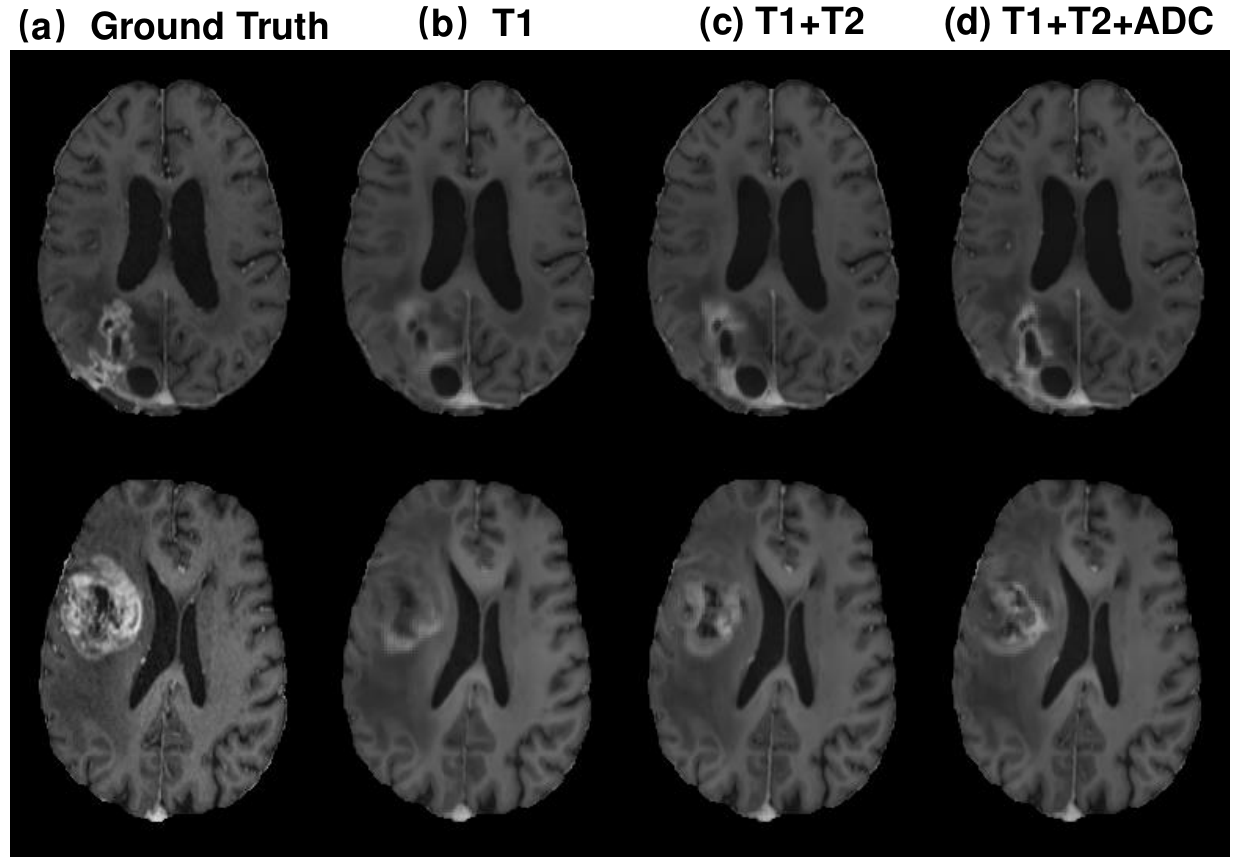}
\end{center}
\caption{Visual performance of training the model by different input modalities. (a) Ground truth images, (b) results by training the model with T1 as input, (c) results by training the model with T1 and T2 as inputs, and (d) results by training the model with T1, T2 and ADC as inputs.}
\label{fig6}
\end{figure}

\begin{figure}[!t]
\begin{center}
\includegraphics[width=1.0\linewidth]{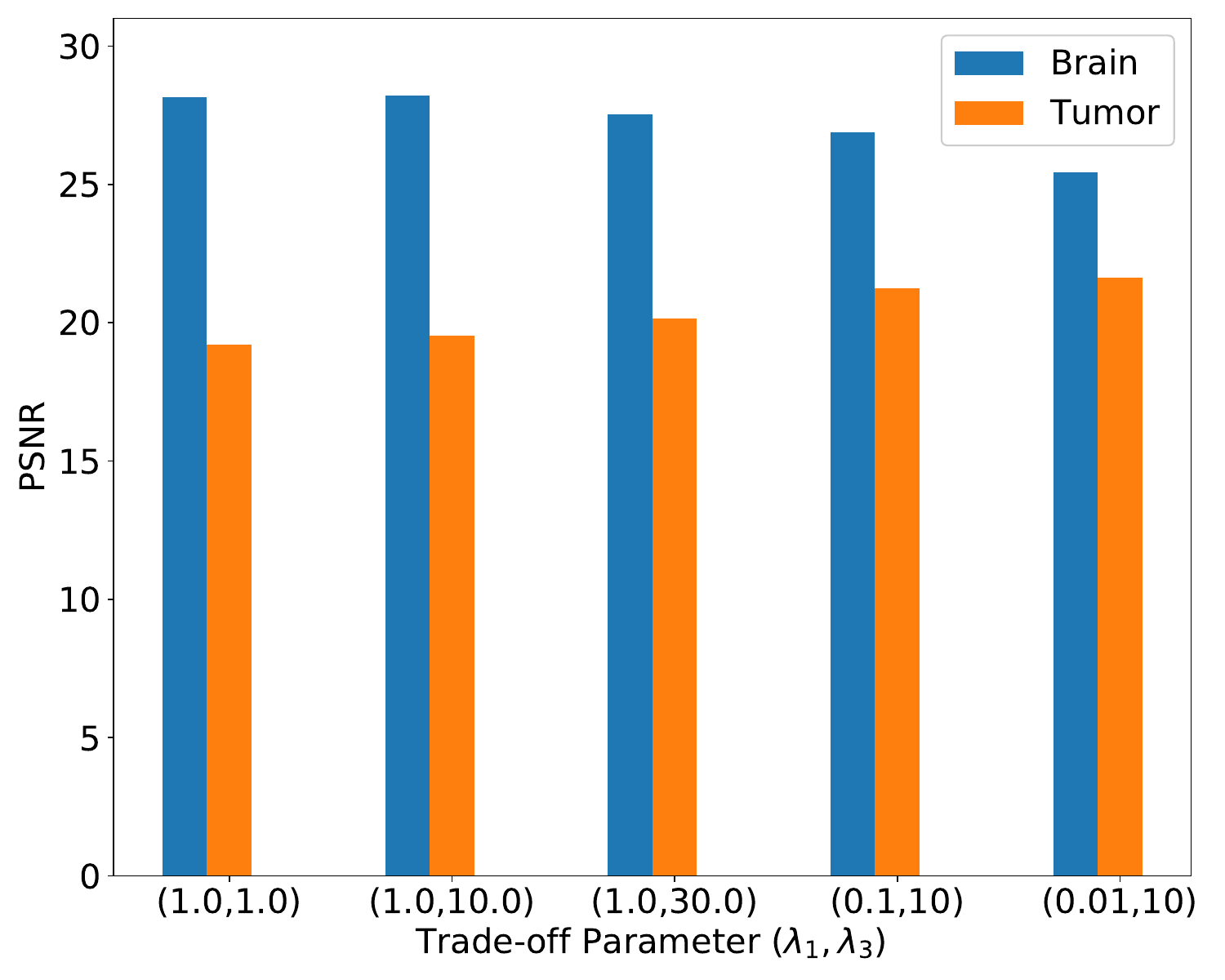}
\end{center}
\caption{Quantitative evaluation of the impact of local loss. The blue bar indicates the PSNR performance on the Brain region of Set A and the orange bar indicates the PSNR performance of the Tumor region of Set B. Note that $\lambda_2$ is set to the same value as $\lambda_1$ throughout the experiments.
}
\label{fig7}
\end{figure}

\textbf{Comparison with State-of-the-Arts}
In addition to the U-Net structure that are used for MRI virtual contrast enhancement\cite{gong2018deep,kleesiek2019can}, we also compare our proposal with several state-of-the-art medical image synthesis methods, including Pix2Pix\cite{isola2017image}, DECNN\cite{xiang2018deep}, LA-GANs\cite{wang20183d} and MedGAN\cite{armanious2020medgan}. Quantitative comparison between our proposal and other state-of-the-art methods is presented in Table \ref{sota}. Pipx2Pix produces the worst result, while DECNN and LA-GANs achieve similar performance and both outperform Pix2Pix. The MedGAN outperforms the previous methods by a large margin, we believe this is because MedGAN utilizes a cascade U-Blocks as generator which is deeper and better designed than the generator in \cite{isola2017image,xiang2018deep}. Compared with the MedGAN, our proposal shows impressive improvement (p-value=0.0064) on brain region, and also significantly outperforms the MedGAN on tumor region by more than 1.3 points  (p-value$<$1e-4), which demonstrate the superiority of the introduced framework. Note that the performance improvement on the tumor region is much significant than the improvement on the whole brain region, this is because the comparison methods does not take into the performance on tumor regions, while we introduce a local loss to improve the performance on tumors. The significant improvement on tumors also demonstrate the effectiveness of the local loss.

\begin{table}[]
\centering
\caption{Quantitative comparison between our proposal and state-of-the-art medical image synthesis methods on test set B}
\label{sota}
\begin{tabular}{|c|c|c|c|}
\hline
\multirow{2}{*}{\textbf{Model}} & \multicolumn{2}{c|}{\textbf{PSNR}} & \textbf{SSIM}  \\ \cline{2-4}
                       & \textbf{Brain}       & \textbf{Tumor}       & \textbf{Brain} \\ \hline
\textbf{Pix2Pix \cite{isola2017image}}                    & 25.90$\pm$1.52             & 18.20$\pm$2.64            & 0.887$\pm$0.025        \\ \hline
\textbf{DECNN\cite{xiang2018deep}}              & 26.48$\pm$1.38             &18.74$\pm$2.31             & 0.898$\pm$0.023      \\ \hline
\textbf{LA-GANs\cite{wang20183d}}              & 26.23$\pm$1.30             &18.69$\pm$2.47             & 0.894$\pm$0.024      \\ \hline
\textbf{MedGAN\cite{armanious2020medgan}}              & 27.04$\pm$1.26             &19.88$\pm$2.42             & 0.901$\pm$0.023       \\ \hline
\textbf{Ours (3D)}              & \textbf{27.62$\pm$1.34}             &\textbf{21.2$\pm$2.36}              &\textbf{0.909$\pm$0.023}       \\ \hline
\end{tabular}
\end{table}

\subsection{Ablation Study}

\textbf{Impact of the Model Architecture}
To demonstrate the effectiveness of the proposed framework, we compare the performance of the final model with four degraded model architectures, which are: (1) Model-A: utilize one FCN module to handle different input modalities, i.e., the three modalities (T1, T2 and ADC) are fused in the input layer; (2) Model-B: remove the 2x downsampling subnetwork and the 4x downsampling subnetwork, only utilizing the high-resolution branch as the FCN generator; (3) Model-C: remove the 1x high resolution subnetwork; (4) Model-D: remove the repeated multi-scale fusion in stage 3 and stage 4. Table \ref{tabMA} shows the qualitative comparison between different model architectures. As can be seen, our proposed model yields notable improvement over the comparison degraded model architectures. In particular, the final model outperforms the Model-A by more than 0.5 points in PSNR, which demonstrates that employ three individual FCN modules for different input modalities is able to preserve the modality-specific information and lead to better performance. The final model outperforms the Model-B by more than 1 point, this is because it is difficult for the model to capture the global information only using the high resolution branch. Besides, The final model also outperforms the Model-C by more than 0.8 points on average, which indicates the effectiveness of maintaining the high resolution representation. Furthermore, the final model also outperforms the degraded Model-D, which proves the effectiveness of using repeated multi-scale fusion stages.

\begin{table}[]
\centering
\caption{Qualitative comparison between the final model and four degraded model architectures.}
\label{tabMA}
\begin{tabular}{|c|c|c|c|}
\hline
\textbf{Model}       & \textbf{A-Brain}    & \textbf{B-Brain}    & \textbf{B-Tumor}    \\ \hline
\textbf{Model-A}     & 27.68$\pm$1.31 & 27.26$\pm$1.28 & 20.63$\pm$2.46 \\ \hline
\textbf{Model-B}     & 26.42$\pm$1.12 & 26.32$\pm$1.20 & 19.54$\pm$2.38 \\ \hline
\textbf{Model-C}     & 27.20$\pm$1.22 & 27.04$\pm$1.19 & 20.39$\pm$2.44 \\ \hline
\textbf{Model-D}       & 27.36$\pm$1.20 & 27.08$\pm$1.18 & 20.68$\pm$2.32 \\ \hline
\textbf{Final-Model} & \textbf{28.24$\pm$1.26} & \textbf{27.62$\pm$1.34} & \textbf{21.20$\pm$2.36} \\ \hline
\end{tabular}
\end{table}

\textbf{Impact of Different Input Modalities} To investigate the influence of different input modalities, we train the model with different sequences. Specifically, in addition to using all three modalities (T1, T2 and ADC) as inputs, we also train the model: (1) using only T1 as input, and (2) using both T1 and T2 modalities as inputs. Experimental results are presented in Table \ref{tab3}. We can conclude from the results that: (1) using T1 alone as input obtains a satisfactory performance; (2) the incorporated T2 and ADC modalities introduced additional information, which further improves performance; and (3) compared to ADC, T2 is more informative and improves performance significantly. Besides, we also present the visual performance of training the model with different input modalities in Fig. \ref{fig6}.

\subsection{Parameter Sensitivity Analysis}
\begin{figure}[!t]
\begin{center}
\includegraphics[width=1.0\linewidth]{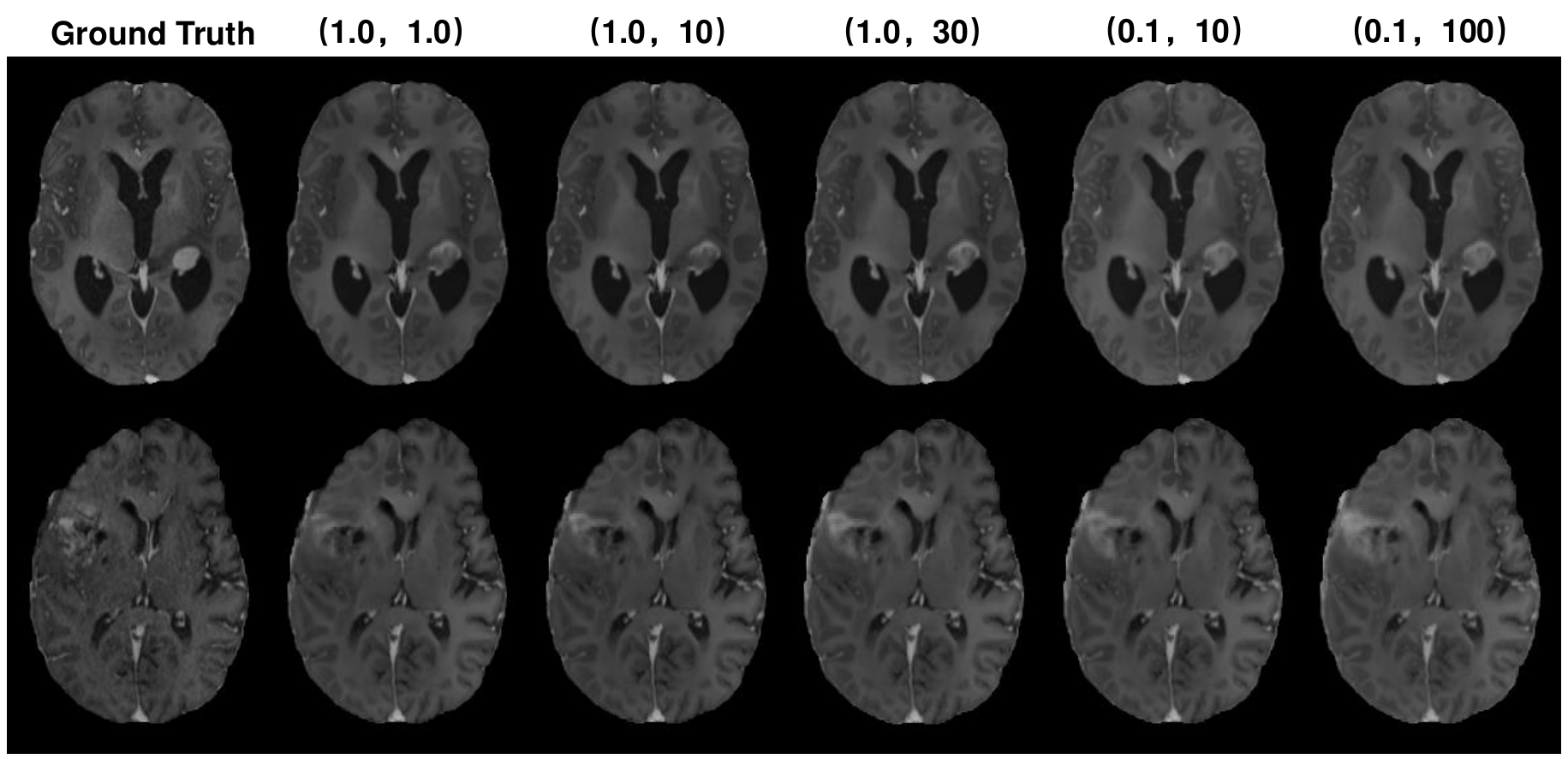}
\end{center}
\caption{Visual performance of training the model by using different trade-off parameters $(\lambda_1,\lambda_3)$. From left to right, the trade-off parameters are set to $(\lambda_1,\lambda_3)=\{(1.0,1.0),(1.0,10),(1.0,30),(0.1,10),(0.01,10)\}$.}
\label{fig8}
\end{figure}

\textbf{Sensitivity of the Local Loss}
In order to investigate the influence of the introduced local loss, we used five groups of representative trade-off parameters to train the model. The parameters $(\lambda_1,\lambda_3)$ were selected from $\{(1.0,1.0),(1.0,10),(1.0,30),(0.1,10),(0.01,10)\}$.  Note that we can not only increase the $\lambda_3$ to improve the influence of the tumor regions as it will cause the gradient to be too large and the network will not converge. Therefore, we promote the influence of the local loss by increasing a ratio $r=\frac{\lambda_3}{\lambda_1}$. $\lambda_2$ is set to the same value as $\lambda_1$ throughout the experiments. Results are presented in Fig. \ref{fig7} and indicate that by increasing the ratio $r=\frac{\lambda_3}{\lambda_1}$, the PSNR performance of the brain region decreased and the PSNR performance of the tumor region increased. We present two representative samples in Fig. \ref{fig8} to demonstrate the influence of local loss visually. By increasing the ratio $\frac{\lambda_3}{\lambda_1}$, tumor enhancement becomes more salient while the overall image becomes more blurry, which is consistent with our quantitative results. The experimental results show that one suitable ratio can be set to $r=\frac{\lambda_3}{\lambda_1}=100$, which leads to better enhancement results for tumors and satisfactory performance for the whole brain.

\begin{figure}[!t]
\begin{center}
\includegraphics[width=0.9\linewidth]{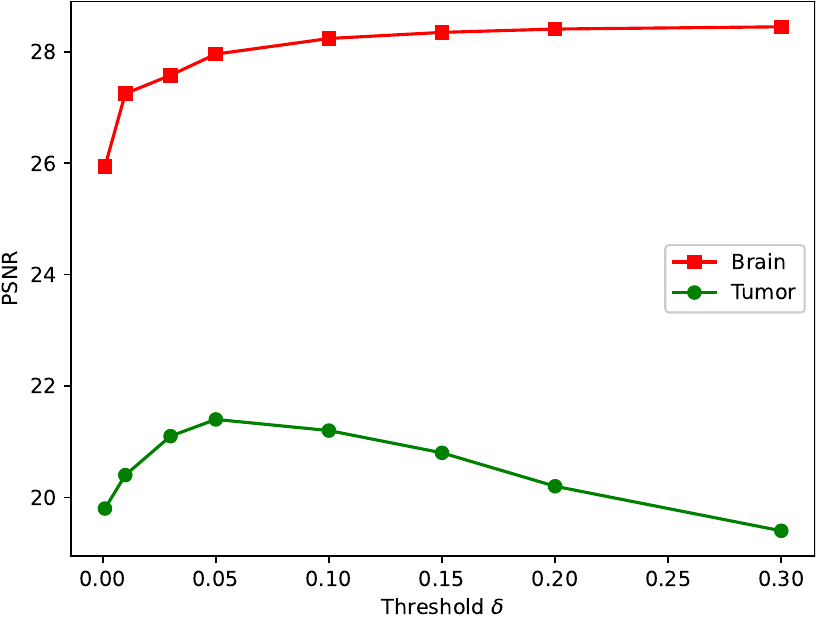}
\end{center}
\caption{PSNR performance on brain region tumor region when threshold is changed from $\delta\in\{0.001, 0.01, 0.03, 0.05, 0.1, 0.15, 0.2, 0.3\}$.}
\label{sen}
\end{figure}

\textbf{Sensitivity of the threshold $\bm{\delta}$}
To demonstrate the robustness of the thresholded mask, we also perform a parameter sensitivity analysis experiment on the thresholds $\delta$. Fig. \ref{sen} shows the variation of PSNR performance on brain and tumor regions when the threshold is changed from $\delta\in\{0.001, 0.01, 0.03, 0.05, 0.1, 0.15, 0.2, 0.3\}$. As can be seen, the PSNR performance on the brain region increases as the threshold increases and gradually stabilize when $\delta>0.1$. Besides, the PSNR performance on the tumor region increases first and then decrease as $\delta$ increases, and shows a bell-shaped curve. We believe this is because when the threshold is set to $\delta<0.03$, the generated mask $\mathbf{M}$ will include more non-tumor areas, and when $\delta>0.15$, the generated mask $\mathbf{M}$ will underestimate the tumor regions. Therefore, based on the performance on brain and tumor region, the best threshold can be set to $\delta\in[0.03,0.15]$.

\begin{figure*}[!t]
\begin{center}
\includegraphics[width=0.9\linewidth]{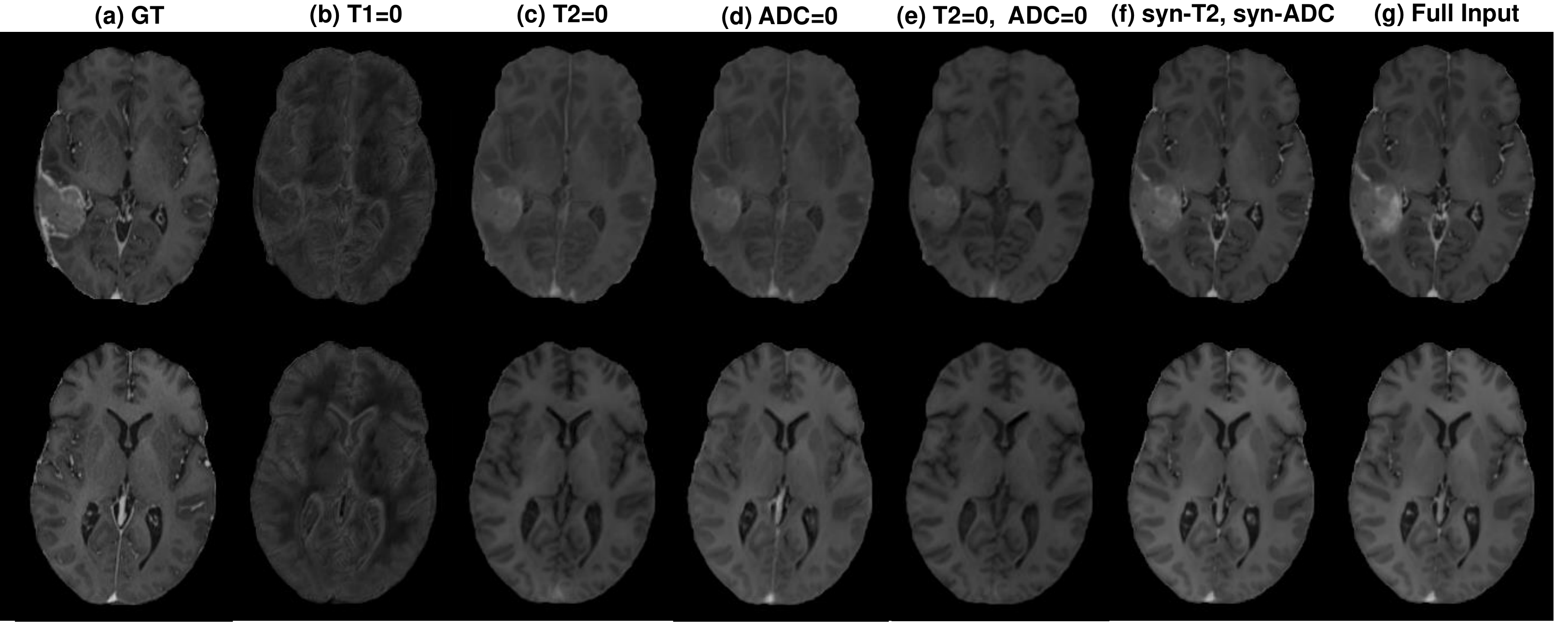}
\end{center}
\caption{Illustration of (a) Group Truth (GT) and model performance when: (b) T1 is unavailable; (c) T2 is unavailable; (d) ADC is unavailable; (e) T2 and ADC are unavailable; (f) utilize T1, synthetic T2 and synthetic ADC as input; (g) utilize full three inputs. Note that we utilize a zero matrix to represent the unavailable input modality during inference.}
\label{missingModality}
\end{figure*}

\subsection{Inference with Missing Modalities}
\begin{figure}[!t]
\begin{center}
\includegraphics[width=0.95\linewidth]{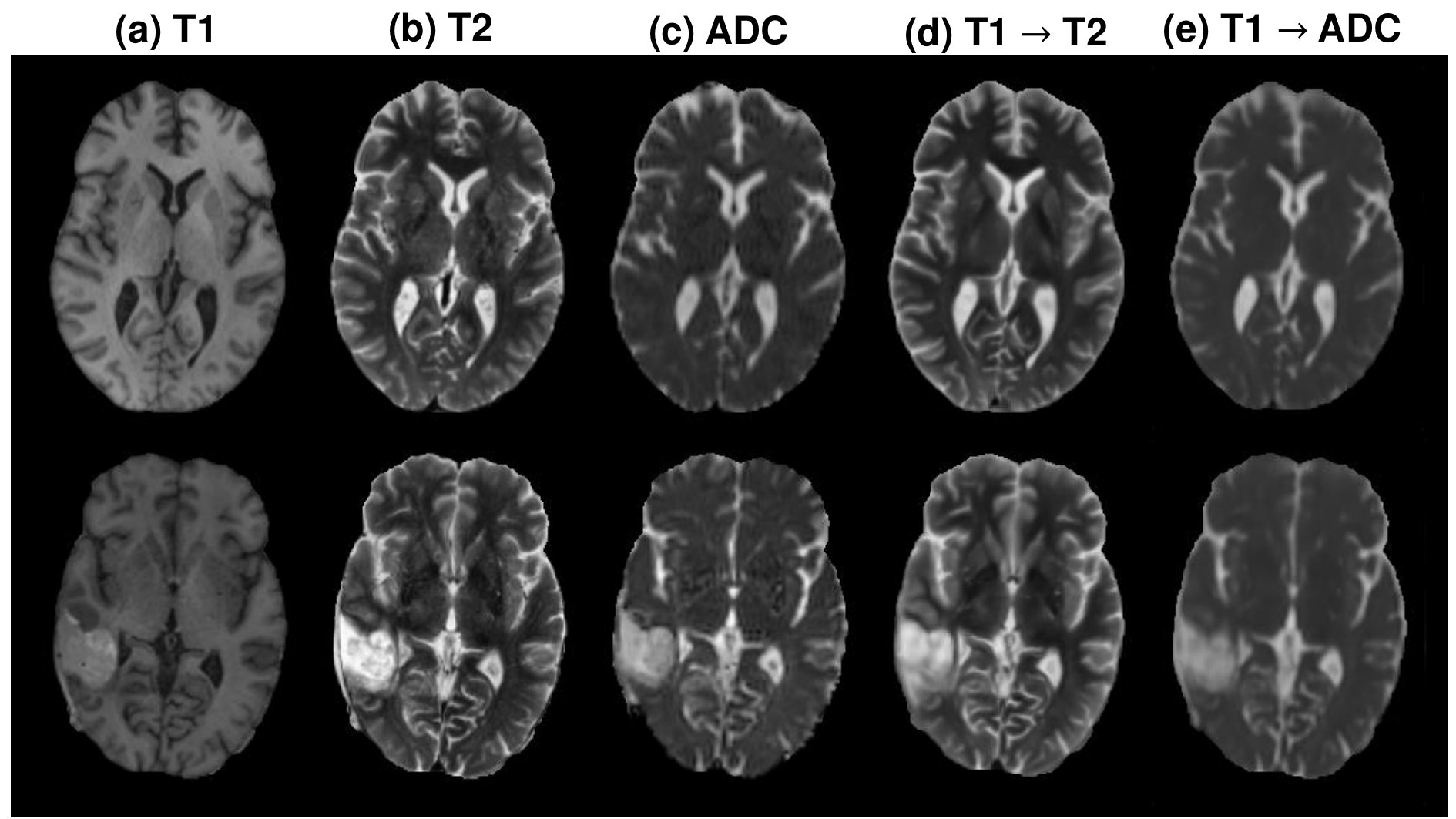}
\end{center}
\caption{The visual performance of the T1$\rightarrow$T2 and T1$\rightarrow$ADC synthesis model. (a) T1; (b) T2; (c) ADC; (d) generated T2 by T1$\rightarrow$T2 synthesis model; (e) generated ADC by T1$\rightarrow$ADC synthesis model.}
\label{imgsys}
\end{figure}

The proposed model works well when all three input modalities are available. However, rather than having complete three modalities, it is common to have missing modalities in clinical scenarios. To understand how the model performs when only a subset of modalities are available, we visualize the inference performance when: (b) T1 is unavailable; (c) T2 is unavailable; (d) ADC is unavailable; (e) both T2 and ADC are unavailable in Fig. \ref{missingModality}. Note that we utilize a zero matrix to represent the unavailable input modality during inference. The results shows that the model performance will be severely degraded when one or two modalities are not available. Besides, compared with the model performance when T2 or ADC is unavailable, the model performs much worse when T1 is unavailable. To ensure that the model is able to produce satisfactory results when only a subset of inputs are available, we also train a T1$\rightarrow$T2 and a T1$\rightarrow$ADC synthesis model utilizing a similar framework. In this way, when T2 or ADC is unavailable, we can first generate the T2 and ADC and then utilize the generated data to synthesize the contrast-enhanced T1 image. Fig. \ref{imgsys} shows the visual performance of our trained T1$\rightarrow$T2 and T1$\rightarrow$ADC synthesis model, which demonstrates that our proposed framework also generates promising results for cross-modal image synthesis task. Fig. \ref{missingModality}(f) illustrates the model prediction of using T1, synthetic T2 and synthetic ADC as input, which shows that using the synthetic input data, our model can generate similar results as using full three inputs.

\section{Discussion}
Several studies have presented methods to generate synthetic contrast. Briefly, Gong et al. \cite{gong2018deep} proposed a 2D U-Net-like model to synthesize the full-dose postcontrast images from precontrast and low dose images. In Kleesiek at al. \cite{kleesiek2019can}, a 3D Bayesian U-Net was utilized to predict contrast enhanced images from 10 multiparametric zero-dose MRI sequences, and Sun et al. \cite{sun2020substituting} proposed a 2D residual attention U-Net to produce contrast in mice brain MR images directly from noncontrast structural images. Compared to these previous studies, our model has the following advantages:
\begin{itemize}
\item In \cite{gong2018deep,sun2020substituting} only one MRI sequence was used for generating full-dose MRI compared to 10 multiparametric MRI scans in \cite{kleesiek2019can}. We believe a single low-dose or noncontrast MRI may miss important information, while the utilization of 10 MRI scans requires a long time to acquire. According to the International Standardized Brain Tumor Imaging Protocol (BTIP) \cite{ellingson2015consensus}, we utilized three informative noncontrast MRI scans (T1, T2 and ADC) for postcontrast MRI synthesis. Our ablation results suggest that using all three of these sequences maximizes performance.

\item To investigate the feasibility of predicting contrast-enhanced MRI sequences from non-contrast or low-contrast MRI sequences, 60 patients are used in \cite{gong2018deep}, 82 patients are used in \cite{sun2020substituting}, and \cite{sun2020substituting} test their idea in mice as a proof of concept. Our study utilized more than 400 patients, allowing us to train a deeper FCN model and obtain state-of-the-art performance.

\item Different from previous methods\cite{gong2018deep,kleesiek2019can} that utilize the off-the-shelf model architectures (2D/3D Unet) and loss function, we present a 3D high-resolution FCN model that maintains high-resolution information throughout the fully convolution stage and aggregates the multi-scale information in parallel. As a result, our model outperforms the 2D/3D U-Net counterparts by more than 1 point in PSNR, and also outperforms several state-of-the-art medical synthesis methods. In addition, we introduced a local loss to improve performance in tumor regions.

\item Previous studies \cite{gong2018deep,kleesiek2019can} obtained imperfect enhancement results in vessels and tumors due to the difficulty of the problem. For example, in Gong et al. \cite{gong2018deep}, enhancement results appear rough in vessels compared to our model. Furthermore, our method achieves a result of 28dB in PSNR on the whole brain region, which is similar to Gong et al. \cite{gong2018deep} despite the fact that our method requires no contrast agent compared to the the low-dose MRI sequences used as input in their work. In terms of PSNR, our model also outperformed Kleesiek et al\cite{kleesiek2019can}. by a large margin even with less input data.

\end{itemize}

While our model demonstrates promising results, there are several limitations.  First, since the tumor regions account for a very small proportion of the entire MRI sequences, the performance on these regions remains sub-optimal. As we do not have precise tumor masks for training, the introduced local loss that we used to balance the contribution of the tumors can be improved. Providing precise tumor masks for the local loss during training will likely improve performance further and is in our future work. In addition, some advanced methods in data imbalance learning or long-tailed distribution learning, such as BBN \cite{zhou2020bbn}, can be introduced to balance the performance of the whole brain region and tumor region. Second, while our cohort was larger than any previously published cohort for the task, performance can likely be further improved by including more training patients, especially a large number of abnormal patients with high diversity. Beyond increasing and diversifying the dataset, future work will also investigate recent advancements in data augmentation.

\section{Conclusion}
In conclusion, the objective of this investigation was to formulate and implement a deep learning model to generate contrast-enhanced MRI sequences from non-contrast MRI sequences, which was expected to eliminate the risk of gadolinium deposition during standard-of-care for brain tumor patients. For this purpose, the largest dataset for the task of MRI virtual contrast enhancement was explored, and a novel high resolution 3D FCN model was designed, which showed superior performance than the counterparts. Besides, we also introduced a local loss to re-balance the contribution of the tumor regions, which leaded to improved performance on tumors. We demonstrate promising  visual and numerical results and obtain state-of-the-art performance. The results suggest great potential in substituting the GBCAs with deep learning to obtain the contrast information in brain MRI. Future work will focus on defective performance on abnormal regions.

\bibliographystyle{IEEEtran}
\bibliography{IEEEabrv,paper}
\end{document}